\documentclass[aps,prl,twocolumn,superscriptaddress]{revtex4-1}
\usepackage{amssymb}
\usepackage{graphicx}
\usepackage{amsmath}
\usepackage{natbib}
\usepackage{bm}

\usepackage{color}

\def\bk{{\bf k}}

\def\bq{{\bf q}}
\def\br{{\bf r}}

\def\la{\langle}
\def\ra{\rangle}

\def\calB{\mathcal{B}}

\def\calA{\mathcal{A}}

\def\e{\epsilon}

\def\pa{\partial}
\def\nn{\nonumber}

\begin{document}

\title{Emergent Symmetry at Superradiance Transition of a Bose Condensate in Two Crossed Beam Cavities}
\author{Zhigang Wu}
\affiliation{Institute for Advanced Study, Tsinghua University, Beijing, 100084, China}
\author{Yu Chen}
\affiliation{Department of Physics, Capital Normal University, Beijing, 100048, China}
\author{Hui Zhai}
\affiliation{Institute for Advanced Study, Tsinghua University, Beijing, 100084, China}
\affiliation{Collaborative Innovation Center of Quantum Matter, Beijing, 100084, China}

\date{\today }

\begin{abstract}

Recently the ETH group has reported an experiment on superradiant transition of a Bose condensate in two crossed beam cavities. The surprise is that they find that across the superradiant transition, the cavity light can be emitted in any superposition of these two cavity modes. This indicates an additional $U(1)$ symmetry that does not exist in the full Hamiltonian. In this letter we show that this symmetry is an emergent symmetry in the vicinity of the phase transition. We identify all the necessary conditions that are required for this emergent $U(1)$ symmetry and show that the ETH experiment is a special case that satisfies these conditions. We further show that the superradiant transition in this system can also be driven to a first order one when the system is tuned away from the point having the emergent symmetry. 

\end{abstract}

\maketitle

Symmetry plays a fundamentally important role in physics. The term of emergent symmetry refers to situations where the symmetry group for the low-energy physics of a quantum system can actually be larger than that of the full Hamiltonian. This usually requires fine tuning of certain parameters in the Hamiltonian. For instance, there are at least two known examples of emergent symmetry in ultracold atomic systems. The first is the emergent Lorentz symmetry in the Bose-Hubbard model, when the system is fine tuned to the particle-hole symmetric point and is in the vicinity of the superfluid to Mott-insulator transition\cite{Sachdev, Auerbach02, Halperin05}. One physical consequence of this emergent Lorentz symmetry is the appearance of the Higgs mode, which has been observed by the Munich group\cite{Bloch12}. Another example is the emergent non-relativistic conformal symmetry in a Fermi gas with a short-ranged interaction, when the interaction potential is fine tuned to a two-body resonance. This emergent conformal symmetry also has important physical consequences, such as the vanishing of the bulk viscosity\cite{Son01,Son05,Son07,Thomas11, Zwierlein11}.  

Putting ultracold atoms into a cavity represents a new hybridized quantum system at the interface of quantum optics and many-body physics\cite{Esslinger07,Colombe,Esslinger13}, and the inevitable decay of the cavity light makes the system intrinsically non-equilibrium.  Previous experiments have studied Bose condensates in a single cavity, and it has been found that the pumping field can drive a superradiance transition, across which the cavity field becomes finite and meanwhile the Bose condensate acquires a density wave order\cite{Esslinger10}. Along this line, experimental and theoretical efforts have also explored such superradiant transitions for strongly interactions\cite{Hemmerich15ex, Esslinger16, Simons09, Simons10, Ciuti13, He13, Hemmerich15th, Yu16, Brennecke16, Mueller16} and with different statistics\cite{ cavityFermion1, cavityFermion2, cavityFermion3, cavityFermion4,TSRExpFermion, Brennecke15,ZhengWei16}. 

\begin{figure}[t]
\includegraphics[width=3.2 in]
{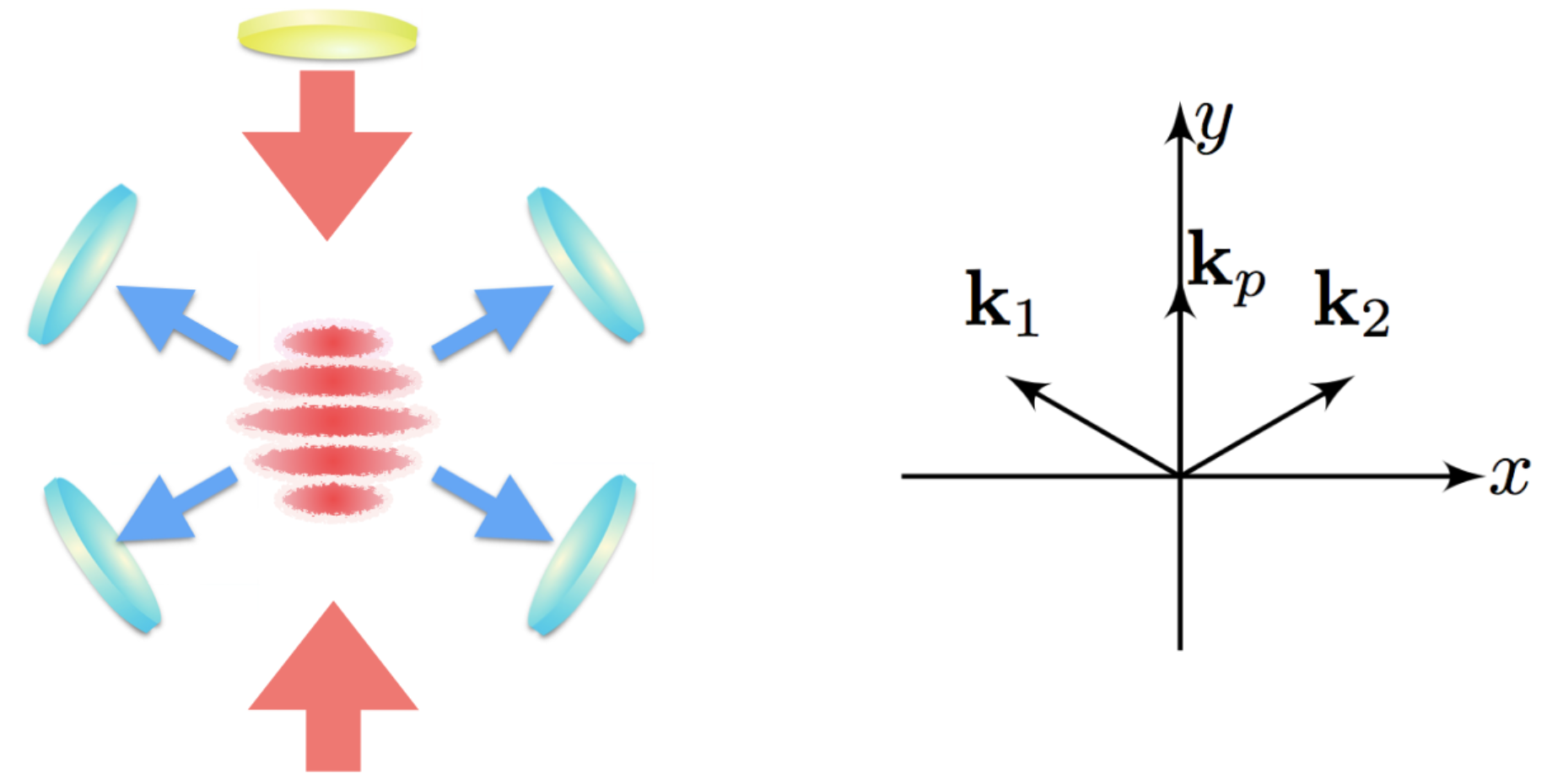}
\caption{Schematic of the two crossed beam cavities (a) and the wave vectors of the pumping field and two cavity fields in ETH experiments (b). }
\label{setup}
\end{figure}

A recent experiment by ETH group has loaded Bose condensate in two crossed beam cavities with same frequency\cite{EsslingerU1}, as shown in Fig. \ref{setup}(a). The pumping beam has a wave vector ${\bf k}_\text{p}$ with a phase $\phi_\text{p}$, and the two crossed cavity beams have wave vectors ${\bf k}_1$ and ${\bf k}_2$ and phases $\phi_1$ and $\phi_2$, respectively. It is quite easy to see that under generic conditions, the system does not have the symmetry of choosing an arbitrary combination of these two cavity modes, because the interference of the pumping beam and different combination of the two cavity beams will lead to different lattice structures, and consequently different energies for the atoms. 

Nevertheless, the recent ETH experiment reveals a surprise. They find that for pumping fields close to the critical value for the superradiant transition, the superradiant light can be emitted to any combination of these two cavity modes when the experiments are repeated under the same conditions. This implies that an additional $U(1)$ symmetry appears aside from the $U(1)$ symmetry breaking in the superradiant transition studied previously. The main point of this letter is to point out that this additional $U(1)$ symmetry is actually an emergent symmetry in the vicinity of the superradiant transition, and we identify that this emergent symmetry requires the following conditions:

(i) The two cavity beams possess mirror symmetry with respect to the pumping beam;

(ii) The wave vectors of the three lasers satisfy ${\bf k}_\text{p}={\bf k}_1+{\bf k}_2$;

(iii) The phases of the three lasers satisfy $\phi_\text{p}-\phi_1-\phi_2=(l_{\phi}-1/2)\pi$, where $l_{\phi}$ is an integer;

(iv) The scattering between the two cavity modes is sufficiently weak so that it can be ignored.  

Without loss of generality, the pumping beam can always be taken along the $\hat{y}$-direction, that is, ${\bf k}_{\text p}=k_{\text p}\hat{y}$. Thus, with conditions (i) and (ii), one can write ${\bf k}_1=k_x\hat{x}+k_{\text p}/2\hat{y}$ and ${\bf k}_2=-k_x\hat{x}+k_{\text p}/2\hat{y}$. In the ETH experiment, the two cavity beams and the pumping beam are aligned in $60$-degree angle with respect to each other, and they have the same wave length. Furthermore the beam phases are chosen to be $\phi_1=\phi_2=0$ and $\phi_{\text p} = \pi/2$. The ETH experiment is thus a special case that satisfies conditions (i), (ii) and (iii). In addition, condition (iv) is also well satisfied in the ETH experiment. That is why the emergent $U(1)$ symmetry is observed there. The following discussion is to show why all these conditions are required for the emergent $U(1)$ symmetry. 

\textit{Model.} In the system of this ETH experiment, a Bose condensate is trapped at the intersection of a pumping beam and two identical high finesse optical cavities. The pumping beam along the $\hat{y}$-direction generates a one-dimensional optical lattice potential $V_{1D}(\br) = ({\Omega_p^2}/{\Delta_a})\cos^2(\bk_{\text p}\cdot \br + \phi_{\text p})$, where $\Omega_p$ is the Rabi frequency and $\Delta_a<0$ is the atom-pump detuning. In the absence of cavity photons, the Hamiltonian for the atomic part is given by
\begin{align}
\hat H_{\text{at}} = \int d\br \hat \psi^\dag(\br)\hat h \hat \psi(\br)+ \frac{1}{2}g_a \int d\br \hat \psi^\dag(\br)\hat \psi^\dag(\br)\hat \psi(\br)\hat\psi(\br), \nn
\label{Hat}
\end{align}
where $\hat h = -\nabla^2/2m + V_{1D}(\br)$ with $m$ being the mass of the atom, $\hat \psi(\br)$ is the atomic field operator and $g_a$ is the interaction strength between atoms. The two cavities, labelled by the index $j=1,2$, are characterized by identical Rabi frequencies $g$ and decay rates $\kappa$. The coupling between the cavity photons and the atoms is given by
\begin{equation}
 \hat H_{\text{ph-at}} = \sum_{j=1,2}\sqrt{UU_{\text p}}\left( \hat a_j + \hat a^\dag_j\right )\int d\br f_j(\br)\hat \rho(\br),
\label{Hphat}
\end{equation}
where $U_{\text p} \equiv \Omega_p^2/\Delta_a$, $U \equiv g^2/\Delta_a$, $\hat a_j$ annihilates a photon in cavity $j$, $\hat \rho(\br) \equiv \hat \psi^\dag(\br)\hat \psi(\br)$ and   
\begin{align}
f_j(\br) \equiv \cos(\bk_j\cdot\br+\phi_j)\cos(\bk_{\text p}\cdot\br+\phi_{\text p}),
\end{align}
which arises from the pump-cavity mode interference. Here we have already used condition (iv) to neglect the lattice potential from the interference of two cavity fields. This can be justified because of the experimental condition $|U |\ll |U_{\text p}|$ and the cavity photon number is small in the vicinity of the superradiant transition. Under the rotating wave approximation, the Hamiltonian for the the entire system can be written as
\begin{align}
\hat H=-\sum_{j=1,2} \Delta_c \hat a_j^\dag \hat a_j+ \hat H_{\text{at}} + \hat H_{\text{ph-at}}
\label{H},
\end{align}
where $\Delta_c$ is the effective detuning between cavity and the pumping field.

It can be shown that under a rotation of the two cavity photon modes, i.e. 
\begin{align}
&\hat{a}^\prime_1=\cos\theta\, \hat{a}_1+\sin\theta \,\hat{a}_2, \label{rot1}\\
&\hat{a}^\prime_2=-\sin\theta\, \hat{a}_1+\cos\theta \,\hat{a}_2 \label{rot2},
\end{align}
$f_j$ transforms as
\begin{align}
&f^\prime_1=\cos\theta f_1+\sin\theta f_2 \label{f1}\\
&f^\prime_2=-\sin\theta f_1+\cos\theta f_2 \label{f2}.
\end{align}
In general, $f^\prime_j$ and $f_j$ are not equivalent, and therefore the full Hamiltonian Eq. \ref{H} is not invariant under the transformation Eqs. \ref{rot1}-\ref{rot2}. Nevertheless, the ETH experiment indeed observed that the superradiant light can populate into either $\hat{a}^\prime_1$ or $\hat{a}^\prime_2$ mode with $\theta$ uniformly distributed between $[0,2\pi)$, which implies that the low energy states of the system are invariant under the transformation given by Eqs. \ref{rot1}-\ref{rot2}. This in fact says that Eqs. \ref{rot1}-\ref{rot2} is an emergent symmetry operation of the system. 

\textit{Emergent Symmetry from an Intuitive Picture.} Before presenting an explicit demonstration of this emergent symmetry, we first give an intuitive picture. Assuming the lattice from the pumping beam is sufficiently deep that atoms are localized at the classical minimum $y_m$ of $V_{1D}({\bf r})$, that is, 
\begin{align}k_\text{p} y_m+ \phi_{\text p} = m\pi,
\end{align}
where $m=0,\pm 1, \cdots$. Under the condition (i)-(iii), it is straightforward to show that 
\begin{align}
f^\prime_1(x,y_m)&= \cos\left(k_x x+\phi_1-\frac{\phi_{\text p}}{2}-\frac{m\pi}{2}-(-1)^{m-l_{\phi}}\theta \right ), \nn \\
f^\prime_2(x,y_m)&=\cos\left (-k_x x+\phi_2-\frac{\phi_{\text p}}{2}-\frac{m\pi}{2}+(-1)^{m-l_{\phi}}\theta \right). \nn
\end{align}
Hence, for atoms fixed at $y_m$, $f^\prime_j$ and $f_j$ are essentially equivalent up to a spatial translation along $\hat{x}$. The physical picture is that the pumping field creates a deep one-dimensional lattice along $\hat{y}$, and at the minimum of this one-dimensional lattice, the cavity beams will further create a one-dimensional lattice along $\hat{x}$. The emergent $U(1)$ symmetry is nothing but the spatial translation of this lattice along $\hat{x}$. 

\textit{Emergent Symmetry from the Ginzburg-Landau Theory.} The superradiant transition at zero-temperature can be well captured by a Ginzburg-Landau type theory, where the order parameter is taken as $\alpha_j\equiv \la\hat a_j\ra$ and it becomes non-zero across the superradiant transition. In the superradiance regime an additional 2D lattice potential from the pump-cavity interference is given by $V_{\alpha} (\br)  =  2\sum_{j=1,2} \sqrt{UU_{\text p}} \text{Re} \alpha_j f_j(\br) $. Neglecting photon fluctuations, the properties of the Bose condensate are determined by $\hat H_{\text{at}} + \hat H_\alpha$, where $ \hat H_\alpha = \int d\br V_\alpha(\br) \hat \rho(\br)$. The boson field satisfies 
\begin{equation}
\left [ {-\nabla^2}/{2m}+V_{2D}(\br) \right ] \hat{\psi} + g_a  \hat{\psi^\dag}\hat{\psi}\hat{\psi}= \mu \hat{\psi},
\label{Psi0}
\end{equation}
where $\mu$ is the chemical potential and $V_{2D}(\br) \equiv V_{1D}(\br) + V_\alpha(\br)$. It is clear that the many-body wave function for the bosons depends on the cavity field strength $\alpha_j$. The latter can be obtained from the equation of motion for the photon field. Taking into account the finite decay rates for the cavities, the equation of motion for the $j$-th cavity photon field is given by
\begin{align}
i\frac{\pa }{\pa t} \hat a_j(t) & = [\hat a_j (t), \hat H ] - i\kappa \hat a_j(t).
\label{EOM_photon}
\end{align}
We are interested in the steady state specified by the condition
${\pa }  \alpha_j(t) /{\pa t}\equiv {\pa }\la\hat a_j(t)\ra/{\pa t} = 0$.
Substituting Eq. \ref{EOM_photon} into this condition and introducing the parameter $\lambda  = -\sqrt{U U_{\text p}} \Delta_c /(\Delta_c^2+\kappa^2)$, one finds that $\text{Im} \alpha_j = -(\kappa /\Delta_c)\text{Re} \alpha_j$ and 
\begin{align}
\text{Re}\alpha_j +\lambda \mathcal{F}_j = 0,
\label{Realphaequ2}
\end{align}
where $\mathcal{F}_{j}  = \int d\br f_j(\br) \langle\psi^\dag(\br)\psi({\bf r})\rangle$. 
Solving Eqs.~(\ref{Psi0}) and (\ref{Realphaequ2}) self-consistently determines the state for both atoms and the cavity fields.

When the photon number is small in the vicinity of the superradiance transition, we can treat $V_\alpha(\br)$ as a perturbation and evaluate the properties of the boson wave function perturbatively in powers of $\text{Re}\alpha_j$. Let us first use the condition that two cavities are mirror symmetric with respect to the pumping beam. In this case $\mathcal{F}_{j}$ can be expanded as
\begin{align}
\mathcal{F}_{j}= c^{(1)}\text{Re}\alpha_j + c_{a}^{(3)}(\text{Re} \alpha_j)^3 +c_{b}^{(3)}\text{Re} \alpha_j(\text{Re} \alpha_{\bar j})^2 +  \cdots, 
\label{rhojpexp}
\end{align}
where $\bar j\neq j$ and the appearance of odd powers of $\text{Re}\alpha_j$ reflects the conservation of crystal momenta. The expansion coefficients $c^{(1)}$, $c_{a}^{(3)}$ and $c_{b}^{(3)}$ can be determined by Eq. \ref{Psi0}. Substituting the expansion into Eq. \ref{Realphaequ2} and neglecting higher orders of $\text{Re}\alpha_j$, we find that if $c_{a}^{(3)}=c_{b}^{(3)}$, $\text{Re}\alpha_1$ and $\text{Re}\alpha_2$ satisfying 
\begin{align}
 \left(\text{Re}\alpha_1\right)^2+\left(\text{Re}\alpha_2\right)^2 = - \frac{ \lambda^{-1}+ c^{(1)}}{ c_a^{(3)}}
 \label{photon_n1}
\end{align}
 are all solutions to Eq. \ref{Realphaequ2}. That is to say, these solutions form a degenerate $U(1)$ manifold. Otherwise, if $c_{a}^{(3)}\neq c_{b}^{(3)}$, it accepts only one solution 
\begin{align}
\left(\text{Re}\alpha_1\right)^2 =
\left(\text{Re}\alpha_2\right)^2
= -\frac{ \lambda^{-1}+ c^{(1)}}{c_{a}^{(3)}+c_{b}^{(3)}}.
\label{photon_n2}
\end{align}
That means that to show the emergent $U(1)$ symmetry is equivalent to showing $c_{a}^{(3)}=c_{b}^{(3)}$.

As said, the coefficients in the expansion of Eq. \ref{rhojpexp} can be determined by Eq. \ref{Psi0}. First of all, for the simplest case of non-interacting bosons at zero-temperature, all bosons are condensed in the ground state of the single particle Hamiltonian $\hat h+V_\alpha({\bf r})$. It is straightforward to calculate the ground state wave function perturbatively in $V_\alpha({\bf r})$, which results in a wave function in powers of $\text{Re}\alpha_j$. Subsequently, by substituting this wave function into the left hand side of Eq. \ref{rhojpexp}, one obtains the coefficients $c^{(1)}$, $c^{(3)}_a$ and $c^{(3)}_b$. Following this route, we obtain that (setting $\e_{00} = 0$)
 \begin{align}
 c^{(1)} &=-2N \sqrt{UU_{\text p}} \sum_{n } {\left|V^{0n}_{0\bq_{1}}\right|^2}/{\e_{n\bq_{1}}}\nonumber\\
 &=-2N \sqrt{UU_{\text p}} \sum_{n } {\left|W^{0n}_{0\bq_{2}}\right|^2}/{\e_{n\bq_{2}}},
 \label{cof_1}
\end{align}
where  $N$ is the number of atoms, $\bq_{j} \equiv  \bk_j - \bk_{\text p}$, $V_{\bar\bk\bar\bk'}^{mn} \equiv \la m \bar\bk| f_{1}| n\bar\bk' \ra$ and $W_{\bar\bk\bar\bk'}^{mn} \equiv \la m \bar\bk| f_{2}| n\bar\bk' \ra$. Here we denote the eigenstates of the single particle Hamiltonian $\hat h$ by $|n\bar\bk\ra$ and their corresponding energy by $\e_{n\bar\bk}$, where $n=0,1,2\cdots$ is the band index and $\bar\bk $ denotes momentum in the one-dimensional Brillouin zone. 

To show the emergent $U(1)$ symmetry, we need the difference between the third order coefficients $c_b^{(3)} - c_a^{(3)}$. Because of the umklapp processes, the expressions for this quantity are different for $\bk_{\text p}\neq \bk_1+\bk_2$ and $\bk_{\text p}= \bk_1+\bk_2$, and they are respectively given by
\begin{widetext}
\begin{equation}
c_b^{(3)} - c_a^{(3)} =  \left\{ \begin{array}{cc}
4N (UU_{\text p})^{\frac 3 2}\sum\limits_n \left ( \frac{|\calA^{0n}_{0,{2\bq_{1}}}|^2 }{\e_{n,{2\bq_{1}}}}-\left [1+(-1)^n  \right ]\frac{|\calB^{0n}_{0,{\bq_{1}-\bq_{2}}}|^2 }{\e_{n,{\bq_{1}-\bq_{2}}}} - 4 \frac{|\calB^{0n}_{0,{\bq_{1}+\bq_{2}}}|^2 }{\e_{n,{\bq_{1}+\bq_{2}}}} \right ); &\bk_{\text p}\neq \bk_1+\bk_2 \\
4N (UU_{\text p})^{\frac 3 2}\sum\limits_n \left ( \chi_{n} \frac{|\calA^{0n}_{0,{2\bq_{1}}}|^2 }{\e_{n,{2\bq_{1}}}}  -    \left [1+(-1)^n  \right ]\frac{|\calB^{0n}_{0,{\bq_{1}-\bq_{2}}}|^2 }{\e_{n,{\bq_{1}-\bq_{2}}}} - 2\chi_{n+1} \frac{|\calB^{0n}_{0,{\bq_{1}+\bq_{2}}}|^2 }{\e_{n,{\bq_{1}+\bq_{2}}}} \right ); & \bk_{\text p}=  \bk_1+\bk_2 
\end{array} \right.
\label{cof_3}
\end{equation}
\end{widetext}   
where $\chi_{n} \equiv 1- (-1)^{n}\cos 2(\phi_{\text p}-\phi_1-\phi_2)$ and the matrix elements are
\begin{align} 
\calA^{0m}_{0\bar\bk} = -\sum_{l,\bar\bk' \neq 0}\frac{V^{0l}_{0\bar\bk'}V^{lm}_{\bar\bk' \bar\bk}} {\e_{l\bar\bk'}}, \quad 
\calB^{0m}_{0\bar\bk} = -\sum_{l,\bar\bk' \neq 0} \frac{V^{0l}_{0\bar\bk'}W^{lm}_{\bar\bk' \bar\bk}} {\e_{l\bar\bk'}}. \nn
\end{align}

\begin{figure}[t]
\includegraphics[width=3.3 in]
{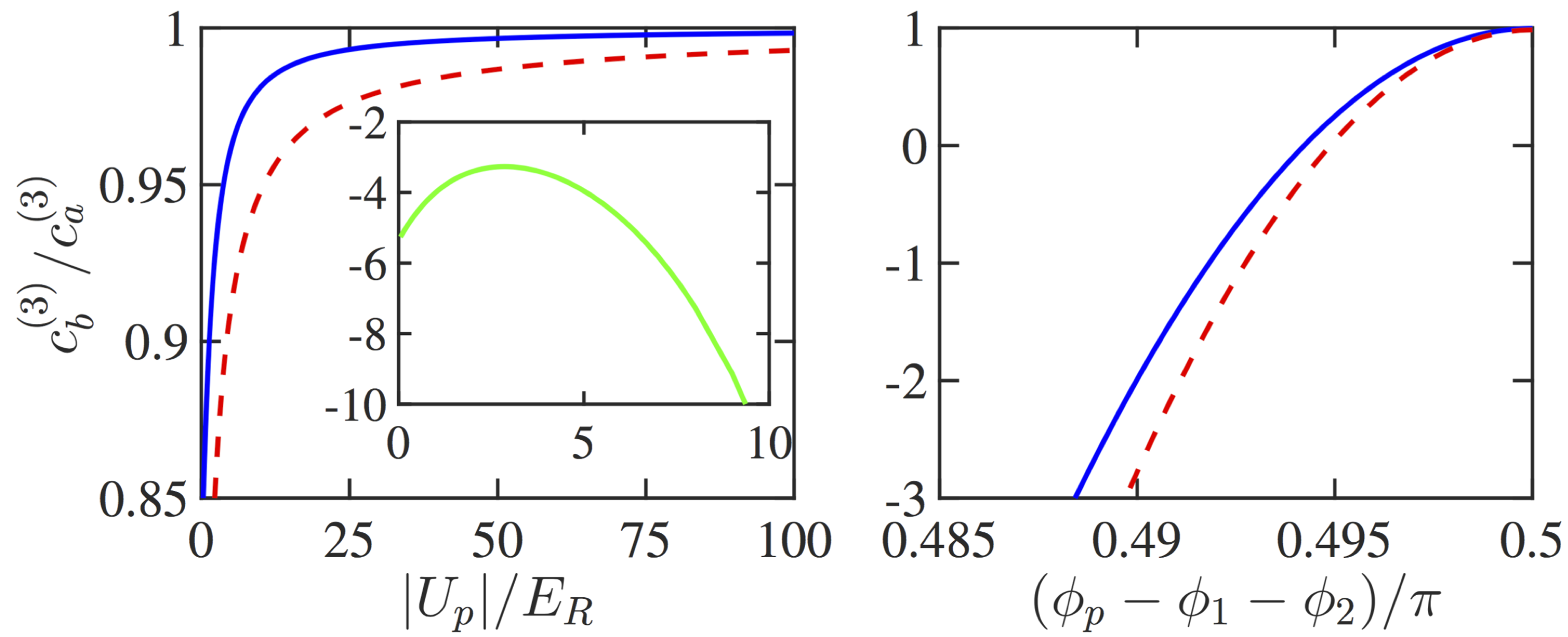}
\caption{$c^{(3)}_b/c^{(3)}_a$ as a function of $|U_\text{p}|/E_\text{R}$ for $\phi_\text{p}-\phi_1-\phi_2 = \pi/2$ (a) and as a function  of $(\phi_\text{p}-\phi_1-\phi_2)/\pi$ for $|U_\text{p}| = 38E_\text{R}$ (b), where $E_\text{R} = k_{\text p}^2/2m$ is the recoil energy. The blue solid and red dashed lines in (a) and (b) are for non-interacting and weakly interacting bosons respectively, with ${\bf k}_\text{p}={\bf k}_1+{\bf k}_2$. For weakly interacting bosons, the calculation is done for $g_a\bar\rho = 0.2$, where $\bar\rho$ is the average atom density. The inset in (a) shows $c^{(3)}_b/c^{(3)}_a$ as a function of $|U_\text{p}|/E_\text{R}$ for non-interacting bosons when the cavities form $45$-degree angle with respect to the pumping beam, i.e., ${\bf k}_1+{\bf k}_2=\sqrt{2}{\bf k}_\text{p}$. }
\label{coeff}
\end{figure}

From Eq.~\ref{cof_3} we can show, under conditions (ii) $\bk_{\text p} =  \bk_1+\bk_2$ and (iii) $\phi_\text{p}-\phi_1-\phi_2=(l_{\phi}-1/2)\pi$, that $\lim_{|U_{\text p}|\rightarrow \infty}c_{b}^{(3)}/c_{a}^{(3)} = 1$, which reveals unequivocally the emergent $U(1)$ symmetry as discussed above. In Fig. \ref{coeff}(a), we consider the ETH experiment set up as an example and show that the ratio $c_{b}^{(3)}/c_{a}^{(3)}$ approaches unity as $|U_{\text p}|$ increases. On the other hand, if $\bk_{\text p}\neq \bk_1+\bk_2$, we see in Fig. \ref{coeff}(a) that $c_{b}^{(3)}/c_{a}^{(3)}$ does not approach unity even if $|U_{\text p}|$ becomes very large. In Fig. \ref{coeff}(b), we also show that $c^{(3)}_b/c^{(3)}_a$ deviates from unity when $\phi_\text{p}-\phi_1-\phi_2$ deviates from $(l_\phi-1/2)\pi$. Hence, together with condition (i) used in formulating these formulae and condition (iv) already used in constructing the model, we have illustrated that condition (i)-(iv) are all needed for the emergent $U(1)$ symmetry.  

The same calculation can be done for a weakly interacting Bose condensate using the Bogoliubov theory. We first obtain the ground state and all excited states with the Bogoliubov theory for Eq. \ref{Psi0} with $\text{Re}\alpha_j=0$. Then we turn on $V_\alpha(\br)$ as a perturbation to obtain the many-body ground state in power of $\text{Re}\alpha_j$, and subsequently determine $c_{b}^{(3)}/c_{a}^{(3)}$. The results are shown in Fig. \ref{coeff} by the red dashed lines, and the conclusion is the same as the non-interacting case. 

\begin{figure}[t]
\includegraphics[width=3.3 in]
{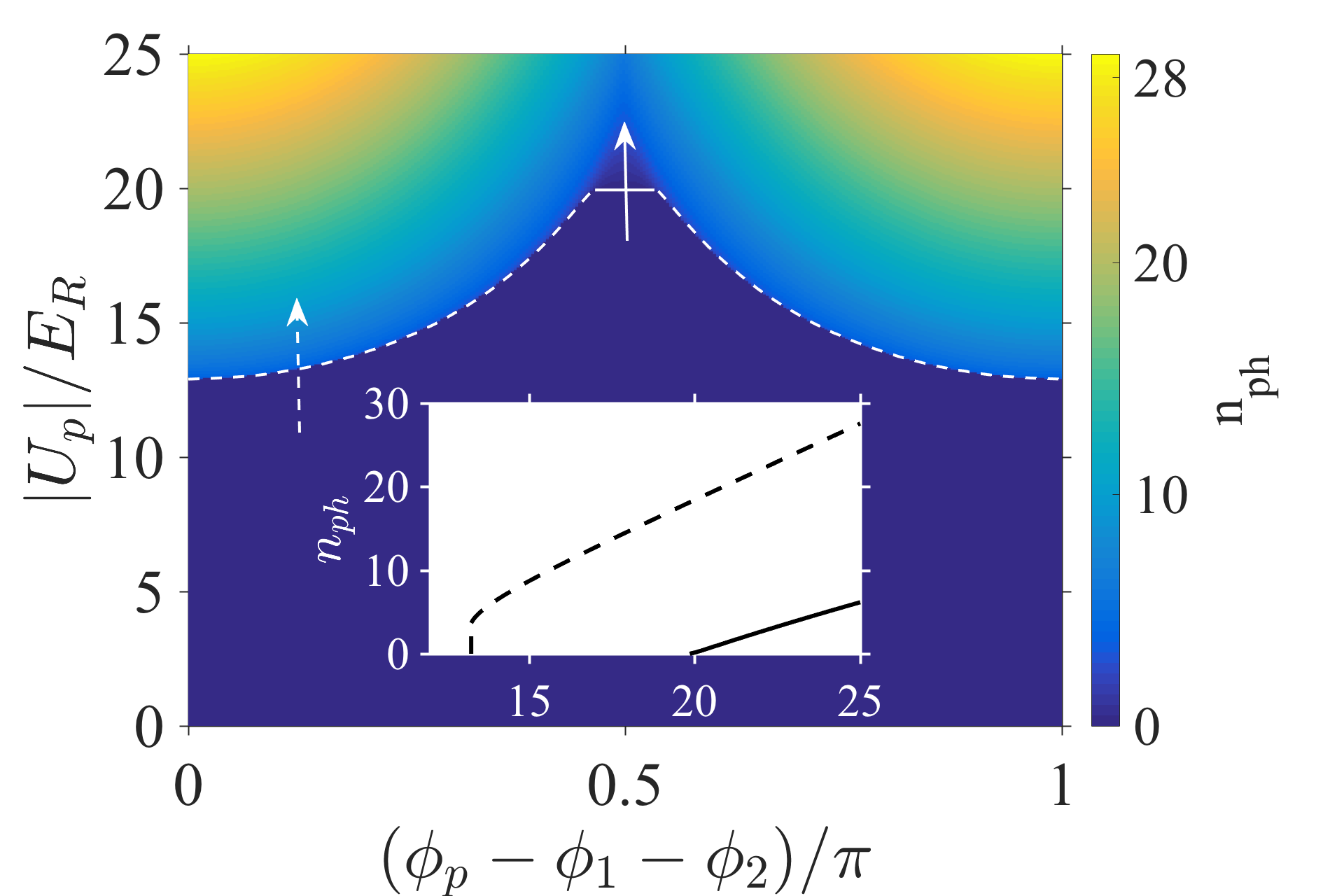}
\caption{The phase boundary between non-superradiant and superradiant phases as functions of the pumping strength $U_\text{p}/E_\text{R}$ and the relative phase $(\phi_\text{p}-\phi_1-\phi_2)/\pi$. The color bar indicates cavity photon number $n_{ph} =|\alpha_1|^2+|\alpha_2|^2$. The dashed line is a first order phase boundary and the solid line is a second order one. The inset shows the photon number as a function of the pumping strength across the first and second order phase transitions. The calculations here are done for the following parameters which are typical in experiments: $N =3\times10^4$, $|U|=4\times 10^{-4}E_{\text R}$, $\Delta_c = 2\pi\times 2\, {\rm MHz}$ and $\kappa = 2\pi\times 150 \,{\rm kHz}$.}
\label{phase}
\end{figure}

\textit{Phase Diagram.} We also notice that in Fig. \ref{coeff}(b) that, when $\phi_\text{p}-\phi_1-\phi_2$ moves away from $(l_\phi-1/2)\pi$, $c_{b}^{(3)}$ can deviate significantly from $c_{a}^{(3)}$, and eventually $|c_{b}^{(3)}|$ becomes larger than $|c_{a}^{(3)}|$, after which the superradiant transition can become a first order one. To determine the first order phase transition boundary, however, we need to solve Eq. \ref{Psi0} and Eq. \ref{Realphaequ2} self-consistently, and the resulting phase diagram for non-interacting bosons is shown in Fig. \ref{phase}. We find that the superradiance phase boundary indeed comprises a second order part  for $\phi_\text{p}-\phi_1-\phi_2$ close to $\pi/2$ and a first order part for $\phi_\text{p}-\phi_1-\phi_2$ away from $\pi/2$. Note that the second order phase boundary is independent of the  relative phase of the optical beams because it is solely determined by the coefficient $c^{(1)}$ which is independent of these phases. 

\textit{Outlook.} In summary, this work explains a unexpected $U(1)$ symmetry discovered in a recent ETH experiment on superradiant of bosons in two crossed beam cavities. We identify this symmetry as an emergent symmetry and determine all the necessary conditions for its appearance. This progress presents an example of emergent symmetry in a non-equilibrium system and offers promise for studying the physical consequences of emergent symmetry in non-equilibrium dynamics. 

\textit{Acknowledgement.} This work is supported by NSFC Grant No. 11604225(YC) No. 11325418 (HZ), MOST under Grant No. 2016YFA0301600 (HZ) and Foundation of Beijing Education Committees under Grants No. KM201710028004(YC).

\textit{Note Added.} When writing this paper, we became aware of two recent papers on the same system, arXiv: 1707.00017 and arXiv: 1707.03907.

\end{document}